%% file: main.tex
\documentclass{article}
\usepackage{spconf,amsmath,graphicx,hyperref,kotex,mathtools,bm,amssymb,xcolor,booktabs}

\def\x{{\mathbf x}}
\def\L{{\cal L}}
\def\e{{\bm{\epsilon}}}
\def\R{{\mathbb{R}}}
\def\EE{{\mathbb{E}}}
\def\Vp{\textcolor[HTML]{0000FF}{\mathbf{V}}_{p}^{\textcolor[HTML]{0000FF}{+}}}
\def\Vm{\textcolor[HTML]{FF8800}{\mathbf{V}}_{q}^{\textcolor[HTML]{FF8800}{-}}}

\def\yp{\textcolor[HTML]{0000FF}{\mathbf{y}^{+}}}
\def\ym{\textcolor[HTML]{FF8800}{\mathbf{y}^{-}}}

\title{DriftTTS: One-stage Single Step Text-to-Speech via Generative Drifting}
\name{Jaehyun Ahn, Dohyun Kim, Heeseung Kim$^{\dagger}$\thanks{$\dagger$ Corresponding Author.}}
\address{Department of AI, University of Seoul, Seoul, South Korea}

\begin{document}
\maketitle

\begin{abstract}
Text-to-speech models built on score matching and flow matching produce natural speech, but only by calling the decoder many times, which makes synthesis slow. Existing ways to cut the number of calls either distill from a multi-step teacher or add a stage on top of a pretrained model, leaving single-step TTS without a pretrained TTS model unexplored. We question whether pretraining a TTS model is necessary, and propose DriftTTS, which adopts generative drifting, an objective that moves iterative sampling from inference into training. Drifting assumes samples of fixed shape, whereas speech has variable length; we address this issue with segment drifting. On LJSpeech, DriftTTS matches the multi-step quality of Matcha-TTS and RapFlow-TTS in a single decoder call, while being faster. To our knowledge, this is the first application of generative drifting to text-to-speech.
\end{abstract}


\input{sections/1_introduction}
\input{sections/2_Generative_Drifting}
\input{sections/3_DriftTTS}
\input{sections/4_experiments}

\input{sections/5_conclusion}

\newpage
\section{Compliance with Ethical Standards}
The listening test was conducted through an online survey platform (Amazon Mechanical Turk) that obtained participant consent and collected only non-sensitive evaluation data.

\section{Acknowledgments}
This work was supported by the Korea Government (MSIT) through the National Research Foundation of Korea (NRF) (No. RS-2022-NR068754, NO. RS-2026-25601142) and by the supercomputing resources provided by the Urban Big Data and AI Institute of the University of Seoul (\url{http://ubai.uos.ac.kr}). We utilized an LLM to assist with language editing, including grammar and wording, and with the development of the project page.

\bibliographystyle{IEEEbib}
\bibliography{main}

\end{document}

%% file: sections/1_introduction.tex
\section{Introduction}

Generative models have advanced since the introduction of diffusion models \cite{NEURIPS2020_4c5bcfec}. Earlier approaches often suffered from over-smoothing, a limitation diffusion models are widely regarded as having largely overcome. Instead of transporting a prior sample to the data distribution in a single step, they train a model to remove a small amount of noise at a time, and generate data by applying it repeatedly, which stably models continuous distributions. Building on this paradigm of discretized iterative sampling, subsequent work has produced score-based generative models \cite{song2021scorebased}, which estimate the gradient of the log-density of perturbed data distributions.

However, iterative sampling requires a large number of function evaluations (NFE), which slows down inference and has motivated efforts to simplify the sampling process. Flow Matching \cite{lipman2023flow} provides a simulation-free framework for training continuous normalizing flows by directly regressing the vector field that transports prior samples to data. Nevertheless, generation still relies on iterative integration, leaving the number of function evaluations as an important bottleneck. MeanFlow \cite{geng2025mean} addresses this limitation by learning the average velocity between two points in time instead of the instantaneous velocity, enabling effective generation with very few function evaluations. Consistency Flow Matching \cite{yang2025consistency} instead imposes velocity consistency across different starting times toward the same endpoint, encouraging straighter sampling trajectories. These approaches have demonstrated strong performance at low NFE in image generation.

Text-to-speech (TTS), which generates natural speech from text, has largely followed this progression.
Grad-TTS \cite{pmlr-v139-popov21a} introduces score-based generative modeling, while Matcha-TTS \cite{10448291} adopts flow matching. Both achieve high-quality speech synthesis but still require a non-trivial number of sampling steps, typically ten or more NFE. IntMeanFlow \cite{11462627}, a zero-shot TTS model, adapts MeanFlow to speech by learning the average velocity along trajectories generated by a pretrained flow matching model, achieving performance comparable to existing models with a single NFE. RapFlow-TTS \cite{park25b_interspeech} instead adapts consistency flow matching by further training a pretrained Matcha-TTS so that points along its estimated trajectory are mapped toward the same data sample. With only two NFE, it matches or surpasses Matcha-TTS using ten. However, both approaches depend on a pretrained TTS model to provide the underlying generation trajectory, increasing training complexity and computational cost.

In this work, we introduce DriftTTS, a TTS model that achieves competitive performance with a single NFE without relying on a pretrained TTS model. DriftTTS builds on generative drifting \cite{deng2026drifting}, which moves iterative sampling from inference to training. Generative drifting formulates sampling as a pushforward map from the prior to the data distribution and directly learns this map during training. At each training step, the model is updated to minimize the magnitude of the drift from its predicted distribution toward the data distribution. To the best of our knowledge, DriftTTS is the first application of generative drifting to TTS.

Our experiments show that DriftTTS with a single NFE performs on par with existing few-step TTS models while requiring fewer function evaluations and no pretrained TTS model. Ablation studies further show that both the feature extractor and segment-level drifting are necessary: without either component, training leads to loudness divergence or false convergence. Synthesized speech samples are available on our demo page\footnote{\url{https://drifttts.github.io/}}.

%% file: sections/2_Generative_Drifting.tex
\section{Generative Drifting}

This section briefly reviews generative drifting \cite{deng2026drifting}. Let a neural network $f: \R^C \rightarrow \R^D$ take $\e \sim p_{\e}$ (e.g., a standard normal distribution) as input and produce $\x = f(\e)$. The distribution $q$ of $\x$ is then the pushforward distribution

\begin{equation}
    q = f_\# p_{\e}.
\end{equation}

Denoting the model at training step $i$ as $f_i$, we call the residual $\Delta \x_i = \x_{i+1} - \x_i$ of the generated sample the \textit{drift}. A zero drift corresponds to a fixed point of the update. To obtain a computable update direction, generative drifting defines a \textbf{drifting field} $\mathbf{V}_{p,q}$ that determines the drift of a generated sample $\x$. Based on mean shift theory \cite{400568}, the field is constructed from an attraction toward the data distribution $p$ and a corresponding interaction with the model distribution $q$:

\begin{equation}
\Vp(\x) \coloneqq \frac 1 {Z_p(\x)} \EE_{\yp \sim p}[k(\x, \yp) (\yp - \x)],
\end{equation}
\begin{equation}
\Vm(\x) \coloneqq \frac 1 {Z_q(\x)} \EE_{\ym \sim q}[k(\x, \ym) (\ym - \x)].
\end{equation}
Here, $k$ is a kernel that measures the similarity between samples, and $Z_p(\x)$ and $Z_q(\x)$ are normalization terms:

\begin{equation}
Z_p(\x) \coloneqq \EE_{\yp \sim p}[k(\x, \yp)],
\end{equation}
\begin{equation}
Z_q(\x) \coloneqq \EE_{\ym \sim q}[k(\x, \ym)].
\end{equation}

The drifting field combines the two interactions as $\mathbf{V}_{p,q}(\x) \allowbreak = \Vp(\x) - \Vm(\x)$. 
Thus, $\Vp$ attracts $\x$ toward nearby data samples, while subtracting $\Vm$ induces repulsion from nearby generated samples. Substituting the two terms gives

\begin{equation}
    \mathbf{V}_{p,q}(\x) = \frac 1 {Z_p Z_q} \EE_{p,\,q}[k(\x, \yp) k(\x, \ym) (\yp - \ym)],
\end{equation}
Generative drifting uses a Laplace kernel with temperature $\tau$:

\begin{equation}
k_\tau(\x, \mathbf{y}) = \exp \left( -\frac 1 \tau |\x - \mathbf{y}|_2 \right).
\end{equation}

In practice, the expectations are estimated over a minibatch, where $\yp$ and $\ym$ are sampled from the data distribution and the model distribution, respectively. For high-dimensional data, generative drifting computes the field in the feature space of an extractor $\phi(\cdot)$ rather than directly on the raw signal. For $\x=f_\theta(\e)$, the training objective is

\begin{equation} \label{eq:drift-loss}
\L = \EE_{\e \sim p_{\e}} \left[\left| \phi(\x) - \texttt{sg}(\phi(\x) + \mathbf{V}(\phi(\x))) \right|_2^2\right],
\end{equation}
where $\texttt{sg}$ is the stop-gradient operator. Numerically, this objective equals the squared magnitude of the drifting field, while the drifted sample is treated as a fixed target during backpropagation. The model is therefore trained to move its samples toward their drifted counterparts, gradually guiding the pushforward distribution $q$ toward the data distribution $p$.

%% file: sections/3_DriftTTS.tex
\section{DriftTTS}

\subsection{Architecture of DriftTTS}

DriftTTS builds on Matcha-TTS and consists of a text encoder, a duration predictor, and a decoder. The text encoder converts the input text into token-level representations. During training, monotonic alignment search (MAS) \cite{NEURIPS2020_5c3b99e8} aligns these representations with the target mel-spectrogram. Specifically, MAS searches for a monotonic alignment that minimizes the $L^2$ distance between the target mel-spectrogram and the text encoder outputs. The resulting alignment provides the duration of each token, which is used to train the duration predictor. It is also used to construct the prior sample $\mu$ by repeating each token representation according to its duration, such that $\mu$ has the same length as the target mel-spectrogram. Since MAS explicitly searches for an alignment that minimizes the $L^2$ distance, the resulting $\mu$ is already close to the target mel-spectrogram. The text encoder and duration predictor are trained using the prior loss $\L_{\textup{prior}}$ and duration loss $\L_{\textup{duration}}$, respectively.

\begin{figure}[t]
  \centering
  \includegraphics[width=\linewidth]{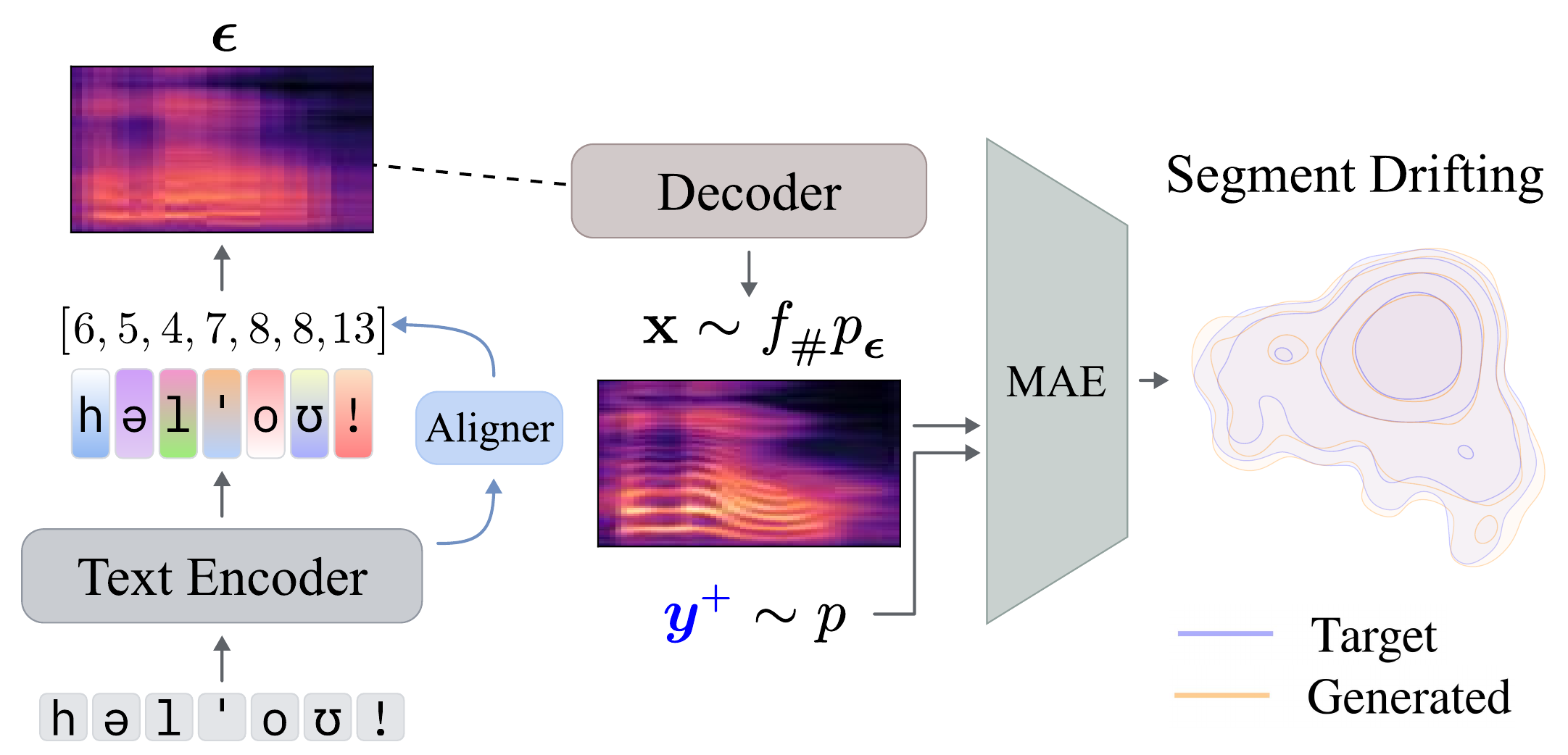}
  \caption{Overview of DriftTTS during training. The text encoder and MAS construct the prior sample $\e=\mu$ by repeating each token representation according to its duration, and the decoder maps it to a generated segment $\x \sim f_\# p_\mu$ in a single call. Segment drifting compares the generated segment with a target segment $\yp \sim p$ in the MAE feature space.}
  \label{fig:overview}
\vspace{-8pt}
\end{figure}

Unlike Matcha-TTS, whose decoder repeatedly predicts a velocity field that is integrated over time, the DriftTTS decoder directly predicts the target mel-spectrogram. Generative drifting does not require a timestep input, allowing us to remove the timestep embedding from the Matcha-TTS decoder and reduce the number of parameters. Fig.~\ref{fig:overview} shows the overall training pipeline of DriftTTS.

\subsection{Segment Drifting}
\label{sec:segment}

Generative drifting is formulated for fixed-shape images with class conditions. TTS, in contrast, involves variable-length speech and text conditions that do not belong to a fixed set of classes. To handle variable length, we sample fixed-length segments from both the prior and target speech during training, following \cite{pmlr-v139-popov21a}. We address the conditioning problem by redefining the starting point, or prior $\e$, from a standard normal sample to the encoder output $\mu$ (i.e., $\e=\mu$). The decoder additionally receives a noise sample as a stochastic condition. Since this prior places each generated sample near its corresponding target, the attraction of $\Vp$ is concentrated around that target, implicitly providing conditional generation.

This choice of prior constrains the positive samples. Since $\mu$ is constructed from an alignment with its target speech, each prior is associated with a specific target rather than being independent of the data sample. Consequently, each generated sample has only one correct destination, and attraction toward another speech segment would provide an incorrect training signal even if the generated distribution matched the data distribution. The corresponding target must therefore be included among the positive samples $\yp$. We satisfy this by using the batch of target segments as $\yp$. In contrast, the negative samples $\ym$ do not require such pairing, and we use the generated segments in the batch, masking out self-interactions when computing $\Vm$.

For the feature extractor $\phi(\cdot)$, we adopt the masked autoencoder (MAE) used in \cite{deng2026drifting}. The MAE is trained to reconstruct masked regions of images. Treating mel-spectrograms as single-channel images, we train the MAE using $4 \times 4$ masks applied with a probability of 50\%. The encoder has a bottleneck structure, and feature representations from corresponding encoder layers are compared in pairs.

\begin{figure}[t]
\centering
\includegraphics[width=\linewidth]{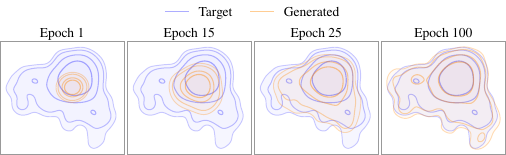}
\caption{Evolution of the generated distribution toward the target distribution during training. Each panel flattens a fixed 80-frame log-mel segment from every evaluation utterance and projects it using a single PCA basis fitted to the targets.}
\label{fig:mel_contour}
\vspace{-8pt}
\end{figure}

Following previous work \cite{deng2026drifting}, we compute the drift at multiple temperatures and MAE encoder layers. Let $i$ index a combination of temperature $\tau$ and MAE layer $l$. We average the losses over $i$ rather than summing them, since summation increases the gradient propagated to the text encoder and reduces the relative influence of $\L_\textup{prior}$, hindering training. The drifting loss $\L_{\textup{drift}}$ is defined as

\begin{equation}
\L_{\textup{drift}} = \EE_{\e,,i} \left[\left| \phi_i(\x) - \texttt{sg}(\phi_i(\x) + \mathbf{V}_{p,,q}(\phi_i(\x))) \right|_2^2\right].
\end{equation}

DriftTTS is trained with $\L_\textup{prior}$ and $\L_\textup{duration}$ for the text encoder and duration predictor, respectively, and with $\L_\textup{drift}$ for the decoder. The final objective is

\begin{equation}
\L = \L_\textup{prior} + \L_\textup{duration} + \L_\textup{drift}.
\end{equation}

%% file: sections/4_experiments.tex
\section{Experiments}

\subsection{System}

We trained both DriftTTS and MAE on LJSpeech \cite{ljspeech17}, following the split used in \cite{10448291}. The predicted mel-spectrogram is converted to speech using HiFi-GAN \cite{NEURIPS2020_c5d73680}, and we therefore follow the mel-spectrogram configuration of HiFi-GAN V1.

DriftTTS was trained for 2000 epochs on four NVIDIA A10 GPUs using AdamW with a learning rate of $1\times10^{-4}$ and a weight decay of 0.01. Gradients are clipped at 5.0, and an exponential moving average (EMA) of the weights with a decay of 0.999 is used for inference. The batch size matters more here than usual: the targets in a batch serve as the positive samples and the generated segments as the negative samples, and previous work reports that generation improves as both increase~\cite{deng2026drifting}. We therefore use the largest batch that fits in memory, 64 pairs of text and speech per GPU. The text encoder is trained on the entire utterance, while the decoder is trained on a segment of 80 frames (${\sim}0.93$,s), and monotonic alignment search runs on GPU with SuperMAS~\cite{supermas}.

The decoder takes a segment of the prior $\mu$ obtained by MAS together with a noise sample as a stochastic condition, and predicts the target segment. The predicted segments serve both as the generated samples $\x$ and as the negative samples $\ym$, and $\Vm$ is computed with the diagonal masked out so that a sample is not repelled by itself. The temperatures are set to $[0.02, 0.05, 0.2]$.

\subsection{Evaluations}

We measure pronunciation accuracy by word error rate (WER), transcribed with \texttt{openai/whisper-large-v3-\allowbreak turbo} \cite{pmlr-v202-radford23a}, perceptual naturalness by a 5-point mean opinion score (MOS) study on the evaluation set, and synthesis speed by the real-time factor (RTF). For each utterance, RTF is computed as the model inference time excluding the vocoder, divided by the generated audio duration. We average RTF over the 500 evaluation utterances at a batch size of one. Every system is run in the same environment on a single GPU, and we report the median over five interleaved runs.

For a controlled comparison, we focus on models that share the Matcha-TTS architecture or are directly built upon it. At a single step, we compare DriftTTS with Matcha-TTS, Matcha-TTS with MeanFlow applied, and RapFlow-TTS. We additionally evaluate the standard multi-step settings of Matcha-TTS and RapFlow-TTS, using ten and two steps, respectively, to examine whether single-step DriftTTS can match their reference generation quality.

We first ablate the MAE by applying segment drifting directly in the mel-spectrogram space, testing whether the feature extractor is necessary. We then replace segment drifting with frame drifting, where each frame is treated independently. This setting is problematic because the prior repeats the same token representation across its duration, while the corresponding target frames are acoustically different. Frame drifting does not preserve which target frame each generated frame should match. As a result, the generated frames can match the overall distribution of target frames and yield $\mathbf{V}=\bm{0}$ even when they are arranged into incorrect speech. We refer to this behavior as false convergence. Following previous work~\cite{deng2026drifting}, we also compare single-temperature losses with the multi-temperature loss. We evaluate these ablations using WER, UTMOSv2, and overall DNSMOS.

\subsection{Results}

\begin{table}[t]
\centering
\caption{Comparison on the LJSpeech evaluation set. MOS is computed from 290 responses of 153 listeners, and RTF is measured on a single NVIDIA A10 GPU.}
\label{tab:main}
\small
\begin{tabular}{lcccc}
\toprule
Model & NFE & WER & MOS & RTF ($10^{-3}$) \\
\midrule
Ground truth       & -- & 1.92 & $4.53\pm0.06$ & -- \\
Vocoded GT         & -- & 1.99 & $4.48\pm0.06$ & -- \\
\midrule
Matcha-TTS         & 10 & 1.93 & $4.38\pm0.08$ & 18.7 \\
RapFlow-TTS        & 2  & 1.82 & $4.42\pm0.08$ & 5.85 \\
\midrule
Matcha-TTS         & 1  & 1.86 & $4.09\pm0.10$ & 4.15 \\
\quad + MeanFlow   & 1  & 1.83 & $4.39\pm0.07$ & 4.19 \\
RapFlow-TTS        & 1  & 1.79 & $4.38\pm0.08$ & 4.19 \\
\textbf{DriftTTS}  & \textbf{1} & 1.80 & $\mathbf{4.43}\pm\mathbf{0.08}$ & \textbf{4.41} \\
\bottomrule
\end{tabular}
\vspace{-8pt}
\end{table}

Table~\ref{tab:main} summarizes the main comparison. DriftTTS achieves comparable WER and MOS to the other synthesized systems. Matcha-TTS loses much of its naturalness at a single NFE. Matcha-TTS with MeanFlow and RapFlow-TTS recover most of this loss, but both require further training of a pretrained Matcha-TTS. DriftTTS reaches this level of quality with a single NFE, trained in a single stage without any pretrained TTS model. It also performs comparably to Matcha-TTS at ten steps and RapFlow-TTS at two steps while running about $4.2\times$ and $1.3\times$ faster, respectively. The single-step baselines remain slightly faster, which we attribute to our decoder implementation rather than the training objective.

Table~\ref{tab:ablation} reports the ablations. The single temperatures perform comparably to the multi-temperature loss in UTMOSv2 and DNSMOS, with no single value consistently outperforming the others. We adopt the multi-temperature loss with $[0.02, 0.05, 0.2]$ from previous work \cite{deng2026drifting}, which also achieves the lowest WER. The feature extractor and drifting unit, in contrast, matter substantially. When segment drifting is applied directly in the mel-spectrogram space without the MAE, every metric collapses as the predicted energy diverges: high-energy mel bins are pushed toward the upper bound of 0, while low-energy bins continue decreasing. We hypothesize that distances between high-dimensional raw mel segments are dominated by coarse energy differences, causing the drifting field to emphasize overall energy rather than local spectral structure. The lower-dimensional frame-wise ablation does not exhibit this divergence, which is consistent with the instability arising from direct kernel computation in the high-dimensional mel space. We leave a rigorous analysis of this behavior to future work.

\begin{table}[t]
\centering
\caption{Ablation on the feature extractor, drifting unit, and temperature. Each row lists its change from the baseline, and the $\tau$ rows use a single temperature in place of the multi-temperature loss. DNSMOS is the overall score.}
\label{tab:ablation}
\small
\begin{tabular}{lccc}
\toprule
Type & WER & UTMOSv2 & DNSMOS \\
\midrule
Baseline          & 1.80 & 3.931 & 3.263 \\
\midrule
w/o MAE           & 20.03 & 0.517 & 1.394 \\
frame, w/o MAE    & 2.72 & 2.137 & 1.724 \\
\midrule
$\tau = 0.02$     & 2.00 & 3.829 & 3.149 \\
$\tau = 0.05$     & 2.00 & 3.982 & 3.244 \\
$\tau = 0.2$      & 1.90 & 3.936 & 3.273 \\
\bottomrule
\end{tabular}
\end{table}

\begin{figure}[t]
\centering
\includegraphics[width=\linewidth]{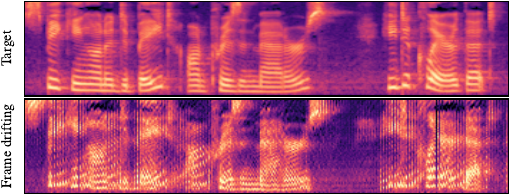}
\caption{False convergence of frame drifting. The target is the mel-spectrogram of a validation utterance, and the frame-wise model predicts it using the alignment found by MAS, so the two panels have the same number of frames.}
\label{fig:frame}
\vspace{-8pt}
\end{figure}

Frame drifting keeps WER reasonable, yet its UTMOSv2 and DNSMOS drop sharply, consistent with false convergence: the prior preserves linguistic content, while frames may match incorrect target frames. As shown in Fig.~\ref{fig:frame}, the prediction follows the structure of the target, but many frames do not connect naturally with their neighbors. Segment drifting avoids this by tying neighboring frames together.

%% file: sections/5_conclusion.tex
\section{Conclusion}

We introduced DriftTTS, a TTS model built upon generative drifting that moves iterative sampling from inference to training, predicting the target mel-spectrogram directly rather than its dynamics, in a single step and without any pretrained TTS model. To apply generative drifting to speech, we proposed segment drifting, which samples fixed-length segments during training and uses the encoded text representation as the prior. It thereby removes the inference cost that grows with the number of steps, while performing on par with the multi-step results of existing models. Extending DriftTTS to multi-speaker and zero-shot synthesis and reducing its dependence on the trained feature extractor remain future work.